# Collaboration and integration through Information Technologies in supply chains

Gilles NEUBERT, Yacine OUZROUT, Abdelaziz BOURAS
[firstname.lastname@univ-lyon2.fr]
Université Lumière Lyon 2
Laboratoire PRISMa – CERRAL
IUT LUMIERE
160 bd de l'universite
69676 BRON FRANCE

**Abstract:** Supply chain management encompasses various processes including various conventional logistics activities, and various other processes These processes are supported – to a certain limit – by coordination and integration mechanisms which are long-term strategies that give competitive advantage through overall supply chain efficiency. Information Technology, by the way of collecting, sharing and gathering data, exchanging information, optimising process through package software, is becoming one of the key developments and success of these collaboration strategies. This paper proposes a study to identify the methods used for collaborative works in the supply chain and focuses on some of its areas, as between a company and its suppliers (i.e., inventory sharing) and its customers (i.e., customer demand, forecasting), and also the integration of product information in the value chain.

## 1. INTRODUCTION

Today business is dominated by an accelerated change in many domains: market, technology, organization, Information, product design, and so on. To face a global economy and rapidly changing markets companies and their suppliers are led to new kind of organization based on value network. To face the globalization of markets the role of manufacturing companies has changed from supplying domestic markets with products, via supplying international markets through export, to supply international markets through local manufacturing [Rudberg & Olhager, 2003].

The complexity of products and technologies is increasing while time to market and product life cycle is decreasing. As mentioned by Momme [Momme 2002], globalization and technological innovation call for improved organizational adaptability and more flexible and advanced systems relative to manufacturing, logistics, engineering, information and process technology. This accelerated environment has driven companies to focus on their core competencies and to outsource some activities or parts of their production. Responsive delivery, without excessive inventory, requires short manufacturing cycle times, reliable processes, and effective integration of disparate functional areas from an inter enterprise point of view as well as from an intra enterprise one.

As highlighted by A.Y. Nahm et al [A.Y. Nahm et al, 2003], manufacturer operating in this so called post-industrial environment focus on customers. Their organizational structure has shift from functional multiple-level hierarchy organization to cross functional orientation, with concurrent information flows and decision making, and relatively few layers in hierarchy. These post industrial organizations focus on more autonomous work, learning, and collaborative management rules.

From many years now, the concepts of Supply Chain (SC) and Supply Chain management (SCM) have been developed in companies' environment, as well as in research papers.

The first one (SC) is relative to the physical organization and flows of the chain from suppliers to end customer. A Supply Chain spans from the raw material suppliers, trough manufacturing, storage, wholesalers and retail companies to the final consumers of the product. It is important to notice that a Supply Chain does not end at the frontiers of an enterprise; it is a chain that comes through different enterprises

The second one (SCM) concerns the organizational aspect and the optimization of the various processes along the Supply Chain. It is mainly based on integration, coordination and information sharing with the goal of achieving a competitive advantage. The important things to notice are that Supply Chain Management has to be seen as a transversal process, and that the objective of the Supply Chain Management is to satisfy the customer at the lowest possible cost

As it is very difficult to have a global view of a Supply Chain and to manage it from suppliers' supplier to customers' customer, Supply chain and Supply chain management can bee seen from an extended point of view as well as from a restrictive point of view. In a broad sense, a supply chain consists of two or more legally separated organizations also named inter-organizational supply chain. The term of intra-organizational Supply chain is applied to a large company with several sites in a narrow sense of the supply Chain (see [Stadtler 2000]).

To support these "intra enterprise" and "inter enterprises" (re)organizations new methods and technologies have arisen. They have to support a better integration from suppliers to customers, a quicker response to customer demand, a better adaptation to market needs and new practices in manufacturing processes. As mentioned by Kobayashi et al [Kobayashi et al, 2003], much attention has been focused on the management methodology of Supply Chain Management (SCM), which integrates business processes from suppliers to customers and manage various tasks, such as sales, manufacturing, logistics, and finance.

In particular, Information Technology (IT), by the way of collecting, sharing and gathering data, exchanging information, optimizing process through package software, has been one of the key developments of SCM. Information Technology makes it possible to process more information, more accurately, more frequently, from more sources, even from all over the globe. And IT makes it possible to digest, to understand, and to act on this growing abundance of information by using sophisticated analysis, modeling, and decision support capabilities [Boyson et al, 2003]

Beyond the goal of improving the performances related to an existing chain for an existing product, Supply chain management is also deployed to improved new product development and launching. Companies are required to develop new product more quickly, less expansive and of better quality. Traditional Research and Development supported by stand alone specification based on product data management is not enough anymore. Customer requirement and suppliers knowledge have to be integrated in the development of new product. This highlights the importance of collaborative design and new kind of information and knowledge are to be shared and exchange along the chain. Product development life cycle naturally includes numerous people, operating in various departments, and typically from multiple companies, each with locations in multiple countries around the world.

Considering all these dimensions of the Supply Chain management, this paper propose a study of the literature to identify the methods used for collaborative works in the supply chain. The selected items are:
- ➢ The relation between a company and its suppliers, such as inventory sharing, etc.
- ➢ The relation between a company and its customers, such as customer demand, forecasting, etc.
- ➢ The relation all along the chain such as information sharing and PLM
- ➢ The optimization of a company processes, such as planning, etc

## 2. INTEGRATION AND INFORMATION TECHNOLOGIES

As the current practice of business is increasingly networked and interconnected, more than ever integration and information are going hand in hand. This last decade, one of the most significant developments in Information Systems has been Enterprise Resource Planning (ERP) systems.

For many companies, ERP is considered as a good answer to replace the fragmented back-office systems and to manage and integrate cross functional business processes. As mentioned by Umble et al (2003), ERP provides two major benefits that do not exist in non-integrated departmental systems: (1) a unified enterprise view of business that encompasses all functions and departments; and (2) an enterprise database where all business transactions are entered, recorded, processed, monitored, and reported. This unified view increases the requirement for, and the extend of, interdepartmental cooperation and coordination.

ERP is standardized software that attempt to put together all departments and function across a company onto a single information system using one unique database. It gets through the old standalone computer software in finance, HR, manufacturing, and so on, and replaces them with a single unified software divided into modules that roughly approximates the old standalone systems. Its role is to provide the connectivity and the common data models needed to link and coordinate the disparate functional silos within the organization as mentioned by White (2002). It aims at integrating key businesses and management processes to provide a consistent view of what is going on in your organization. The great benefit of ERP is integration; as it combines each department program and data all together into a single, integrated software program that runs off a single database, all employees can use the same information and business processes and get the same results when the system is required. In this way the various departments can easily share information and communicate with each other

Implementing an ERP increases drastically the transparency of the information system. People now need to fill in real time the data bases so that each one in the process can use it for its own job. As noticed by Koch (2002) "*People in the warehouse who used to keep inventory in their heads or on scraps of paper now need to put that information online. If they don't, customer service reps will see low inventory levels on their screens and tell customers that their requested item is not in stock*".

For this reason, ERP systems are essentially considered as process-oriented IT tools for improving business performance [Al-Mashari et al, 2003]. They are highly complex information systems that work essentially at integrating data. The commercially available software packages promise seamless integration of all information flows in the company-financial and accounting information, human resource information, supply chain information, and customer information [Umble et al, 2003].

But ERP projects are not only technical projects involving the information system or the computers configuration. One of the major critical success factors highlighted during ERP implementation is the organizational change management. ERP project have to be manage as Business Process Reengineering project that shake up the whole organization. ERP software embedded standardized solution so that it asks people to change the way they do their jobs. Of course most ERP are build with a certain amount of "flexibility" in their solutions but it will nether fit the way of doing business as well as the old specific in-house developments. That's the reason why implementing an ERP in a company usually involves changing business processes; and employee resistance to these changes can be a major brake on the project success. As noticed by Koch (2002), *"If you simply install the software without changing the ways people do their jobs, you may not see any value at all—indeed, the new software could slow you down by simply replacing the old software that everyone knew with new software that no one does"*.

As mentioned by Lamber and Cooper (2000), successful SCM requires a change from managing individual functions to integrating activities into key supply chain processes. ERP Systems are a good response (not the only one) to this organizational need of change inside a company. They can now be considered as the core backbone to which other applications may be linked as a mean of extending the functionality of enterprise software.

ERP packages are incomplete to fit an individual company's business processes. Further on, it is common to see companies choose one ERP vendor for financials while choosing another software vendor for human resources applications or production planning. As firms moved toward business process rather than departmental applications, the need to integrate functionalities is increasing: Shop floor, inventory, accounts receivable and advanced planning application, HR, etc. need to communicate so that companies can make accurate promises to customers, and executives can decide more quickly.

For some enterprises business processes and information are still maintained in customised legacy systems based on dated technologies and stored in standalone databases.

Whether it is legacy systems or a packaged business solution, applications in most cases could not effectively exchange information and the integration challenge resides now in uniting the many different "island of automation" that exist within an enterprise. As written by Erasala et al (2003), the need of the hour is to be able to share information and business processes without having to make sweeping changes to existing applications and data structures. A promising approach to achieve this is Enterprises Application Integration (EAI). EAI represents an attractive proposition to firms, since it offers them the opportunity to leverage their systems into a seamless chain of processes and present a unified view of their business to customers.

EAI is not like Application Program Interfaces (API's) or 'point to point' solution but it creates a common way for both business processes and data to speak to one another across applications.

Irina et al. (2003) propose a taxonomy to help managers to identify technologies that can be used for enterprise and cross-enterprise application, which can lead to the development of an integrated infrastructure that support intra-and inter-organisational applications [Irani et al, 2003]. They distinguish: (1) intra-organisational application integration, (2) inter-organisational application integration (cross enterprise), and (3) hybrid application integration (B2C application)

## 3. COLLABORATION AND IT

Historically, companies have focused only on their resources, constraints, and policies to make decisions and reduce costs [Lambert & Cooper, 2000]. With intense competition and reducing profit margins, this approach is no longer sufficient. They need to consider the interactions with their suppliers and customers and incorporate them into their decision making process. They also need to reformulate their business policies to enable them to incorporate the information regarding their supply chain into their decisions.

As companies outsource more and more of their current in-house activities, they will have to develop tools the software tools to control and collaborate with their outsourcers as well as evaluate their performance in real time. These types of closely tracked relationships amongst supply chain partners will put an increasing level of pressure on collaboration and execution efficiencies [Wilson, 2002]. As mentioned by Akintoye et al., the various definitions which have been proposed indicate that SCM prescribes organizational restructuring, extended to the achievement of a company-wide collaborative culture [Akintoye et al., 2000]. In their paper Thomas and Griffin defined three categories of operational coordination: (1) Buyer-Vendor coordination, (2) Production-Distribution coordination and Inventory-Distribution coordination [Thomas and Griffin, 1996]. They show the breakdown of some models reviewed in the literature according to their typology. From then, the rapid evolution of Information Technology and the increasing use of internet technology have drastically changed our way of doing business. As a consequence, the need for designing new business processes with the assumption that there are no barriers between departments and enterprises has increased. The internet can redefine how back-end operations – product design and development, procurement, production, inventory, distribution, after-sales service support, and even marketing – are conducted, and in the process alter the roles and relationships of various parties, fostering new supply networks, services and business models [Lee and Whang, 2001].

In traditional organization, each company operates individually and manage its own situation creating multiple decision points and minimal information sharing. Each actor is responsible for his own inventory, production and distribution activities. On the other side, Supply Chain Management stresses the need for process integration and new practices to share responsibilities among supply chain partners. Effective supply chain management requires coordination among the various channel members including retailers, manufacturers, and intermediaries. Programs such as vendor managed inventory (VMI), continuous replenishment planning (CRP), and Customer Relationship Management (CRM) have been advocated by some as promising approaches to supply chain coordination. [Waller & al., 1999].

### 3.1 Through existing collaborative processes: from VMI to CPFR

Analyzing the inventory decision in a distribution network has probably been the most extensively research area. Traditionally, the focus has been on understanding the structure of the optimal decision that minimizes a cost function ( in term of ordering quantities or review levels, etc.) assuming some hypothesises (known demand, etc.) (see [Erenguç et al, 1999] ).

In order to reduce lead –time and inventory cost, suppliers and customers have developed cooperative models based on information sharing. As noticed by Huang et al [Huang et al,

2003] sharing of inventory level information can improve decision such as safety stock placement, order replenishment and transhipment in a distribution supply chain. Knowing the inventory level of downstream facilities can reduce the inventory holding cost of the whole supply chain. Such a practice is called Vendor Management Inventory (VMI).

VMI is an operating model in which the supplier takes responsibility for the inventory of its customer. In this kind of partnership, the supplier makes the main inventory replenishment decisions for the customer. By this way, the supplier controls the buyer's inventory level, so as to ensure that predetermined customer service levels are maintained. VMI systems achieve its goal through more accurate sales forecasting methods and more effective distribution of inventory in the supply chain

Achabal et al give an overview of the models developed for the VMI and develop a decision support systems that combines inventory optimization methods an promotional response models [Achabal et al, 2000]. It highlights the role of the collaboration between customer (ie retailer) and supplier (ie vendor), the importance of information sharing between the two partners, and the integration between inventory and forecast models.

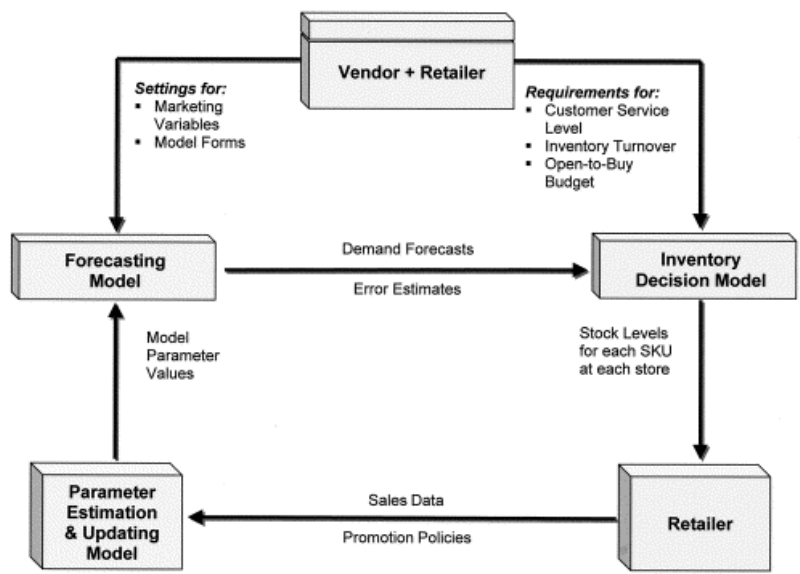

Figure 1: Models used for VMI Decision Support System [Achabal et al, 2000]

Before any supply chain coordinating arrangement is entered into, the retailer and manufacturer manage their inventories independently, with the manufacturer in a less favourable position regarding information. The manufacturer relies on historical order data from the retailer to predict both future ordering patterns of the retailer and true demand patterns of the retailer's customers. Obviously, the manufacturer's understanding of the retailer's actual demand pattern is distorted by the latter's demand forecasting and ordering behaviour. Therefore, it is difficult to closely synchronize the manufacturer's delivery at the retailer's location with demand. A mismatch between delivery and demand also causes fluctuations in order releases. VMI reduces information distortion, which is one cause of the bullwhip effect.

In addition, VMI provides the supplier with the opportunity to better manage its own production, inventory, and transportation costs. For suppliers, the major attraction of VMI is in smoothing demand [Kaipia and Tanskanen, 2003]. Large, infrequent orders from customers force suppliers to maintain inventories that enable them to respond to the uneven demand. In VMI, the supplier is able to smooth the peaks and valleys in the flow of goods, and therefore to keep smaller buffers of capacity and inventory. The impact of demand amplification is dampened as the manufacturer now receives a direct view of end customer demand patterns and can use this in forecasting [Disney et al, 2003].

VMI can increase customer service levels and reduce stock-outs through better understanding of demand and more sophisticated inventory policies. For retailers, benefits come from increasing service levels with lower inventory costs.

The notion underlying such programs as VMI and continuous replenishment planning (CRP is a move from pushing products from inventory holding areas to pulling goods onto grocery shelves based on consumer demand) has been around for a long time, but recent advances in technology, including electronic data interchange and the Internet, are now making these programs feasible. For instance, two major barriers associated with managing another organization's inventory at remote locations – knowing the exact inventory status and predicting the demand activity – are being overcome as a result of the real-time capture and transmission of transaction information ([Stank et al., 1999](Stank et al., 1999)).

Companies are now evolving to collaborative replenishment and co-managed inventory solutions which are more balanced approach to CRP initiatives. In collaborative replenishment, both the customer and the supplier have the visibility to demand, and longer-term forecast are used to highlight upcoming surges and lulls that call for advanced planning.

This is one issue of Collaborative Planning, Forecasting and Replenishment (CPFR), which is a triptych of collaborative planning, collaborative forecasting and collaborative replenishment. This model was proposed by the Voluntary Inter-Industry Commerce Standards Organisation (VICS – see [http://www.cpfr.org](http://www.cpfr.org) ). CPFR is a process model, shared by a buyer and supplier, through which inventory-status, forecast-, and promotion oriented information are shared and replenishment decisions are made. It helps trading partners generate the most accurate forecasts possible and set highly effective replenishment plans. CPFR is centred on open business relationships that share information, data and communication; it requires information technology to build, share, and adjust on-line forecasts and plans.

CPFR is the first step toward developing a broader collaborative relationship between retailers and suppliers. Collaboration is a long-term strategy that give competitive advantage through overall supply chain efficiency. CPFR streamlines the flow of physical goods to reduce distribution costs and supply chain inventory and improve in-stock positions and sales.

It also facilitates a number of intangible benefits between trading partners and suppliers including improved trust, increased collaboration in areas outside the scope of CPFR, and idea sharing.

The two tables below show the typical CPFR Benefits for retailers and manufacturers:

| Retailer Benefits | Typical Improvement |
|---|---|
| Better Store Shelf Stock Rates | 2% to 8% |
| Lower Inventory Levels | 10% to 40% |
| Higher Sales | 5% to 20% |
| Lower Logistics Costs | 3% to 4% |

| Manufacturer Benefits | Typical Improvement |
|---|---|
| Lower Inventory Levels | 10% to 40% |
| Faster Replenishment Cycles | 12% to 30% |
| Higher Sales | 2% to 10% |
| Better Customer service | 5% to 10% |

Source: AMR Research April 2001- [AMR 2001]

## 3.2 Beyond industrial application: AI and Collaborative Decision Making

As previously mentioned in this paper SCM encompasses various processes and is mainly supported by coordination and integration mechanism. These processes include of course conventional logistics activities, such as inventory management, transportation, order processing, and other processes such as customer and supplier relationship management, product development, and so on.

Decisions are usually based on department's own constraints and are optimized locally within the departments but do not assure a global optimum for the enterprise. Of course, decision support tools exist for local decision-making, e.g. planning and scheduling systems, inventory management systems, market trading optimization systems, etc. An integration of these tools would not solve the problem and there is a need for a unified approach for modelling and analysis of supply chains, which explicitly captures the interactions among enterprises and within the departments **(Julka & al. 2002).**

Many information technologies have been developed for handling transactions in supply chains such as electronic document interchange (EDI) and enterprise resource planning (ERP). Lately, Internet-based technologies such as the ebXML and Web Service (**Walsh, 2002**) are emerging. However, despite the merits of these technologies, there are limitations in the flexibility and dynamic coordination of distributed participants in supply chains. The agent system is an alternative technology for supply chain management because of the features such as distributed collaboration, autonomy, and intelligence (**Nissen, 2001; Yuan et al., 2001**)

### 3.2.1 Multi-Agent Systems technology

Agents are considered as entities with goals, capable of actions endowed with domain knowledge and situated in an environment to achieve internal goals, an agent must reason about its environment, to generate plans and execute them. *Multi Agent System* (MAS) is an organization of agents, interacting together to collectively achieve their goals **(Chehbi & al. 2003)**. Software agents can be defined in different ways depending on the way they are implemented and the tasks they perform. **(Wooldridge & al.,1998)**.

Many researchers in distributed artificial intelligence have dealt with developing theories and techniques to study and apply multi agent systems in organizations management where agents collaborating will play a major role. A good summary of some applications based on SMA can be found in **(Wu, & al., 2000).**

Research literature on intelligent agent system architectures has proven that problems that are inherently distributed or require the synergy of a number of distributed elements for their solution can be efficiently implemented as a multi-agent system (MAS) **(**Jennings, Sycara, & Woolridge, 1998; Ferber, 1999**).** All agents participating in MAS communicate with each other by exchanging messages, encoded in a specific Agent Communication Language (ACL). Each agent in the MAS is designated to manipulate the content of the incoming messages and take specific actions/decisions that conform to a particular reasoning mechanism designed by the agent programmer.

Previous researches on agent-based supply chain management can be divided into three types (Ahn 2003):

- **Agent-based architecture for coordination** : in this first type of research, functional units of the chain are modelled as independent and collaborating agent systems, such as logistics agent, transportation agent, scheduling agent, etc. (Fox et al., 2000; Ito & Abadi, 2002). The aim of this type of research is to show how to coordinate various activities of the agents in different conditions. For example, Reis et al. (2001) showed how distributed agents can be coordinated for effective production scheduling.

- **Agent-based simulation of supply chains**: the second type of research shows how agent-based supply chains can gain visibility and efficiency through simulation under various strategies (Fox et al., 2000). Fox et al. (2000) showed how agent systems can be modelled for supply chain simulation when there are unexpected events in some of the companies within the supply chain.

- **Dynamic formation of supply chains by agents**: the last type of research shows how supply chains can be formed dynamically meeting various environmental constraints that may change over time**.** In this type of research, agent systems that represent each company in a supply chain perform negotiations with other agents to form a virtual supply chain, which lasts for a certain period to achieve a common objective. The major concern in this type of research is to show how agent technology enables more flexible organization of supply chains using autonomous, collaborative, and intelligent features. For example, Chen et al. (2000) show how virtual supply chains can be formed by solving distributed constraint satisfaction problems. In Walsh and Wellman's research (Walsh & Wellman, 1999), supply chains are formed by subcontract auctions of participating agents.

Ahn & All (2003) explain that one of the main benefits of using agent technology for supply chain management is the dynamic formation of supply chains using negotiations or contracts by agents (Chen et al., 1999). However, the limitation of existing agent systems is that it is difficult to make agent based supply chains adapt to new products or new trading partners because agent systems usually use a fixed set of transaction sequences, which is a critical barrier to adaptability (Freire & Botelho, 2002). So, the purpose of their research is to propose a flexible agent system, which is adaptable to the dynamic changes of transactions in supply chains.

### 3.2.2 Collaboration and Multi-Agent Systems

In the last few years, multi-agent systems have been a preferred tool for solving supply chain problems:

- Goodwin, Keskinocak, Murthy, Wu, and Akkiraju (1999) present a framework for providing decision support for an online exchange. They use a multi-agent system to find matches of demand and supply on the exchange and provide the user with the best set of transactions..

- Sauter, Parunak, and Goic (1999) present an architecture called Agent Network for Task Scheduling (ANTS), inspired by insect colonies and humans. Agents represent elements in the supply chain and within a factory. Each firm in turn is viewed as a small supply chain, thus the interface between agents within a

- Julka & al. (2002) present a framework which helps to analyze the business policies with respect to different situations arising in the supply chain. This architecture is based on amulti-agent system.

- Swaminathan, Smith, and Sadeh (1998) present a modelling and simulation framework for developing customized decision support tools for supply chain reengineering. Agents represent supply chain entities, e.g. customers, manufacturers, and transportation.

- Another work presents the idea of open tender concept with blackboard-based negotiation to develop a collaborative supply chain system (CSCS) (Ito & Salleh 2000). Blackboard-based negotiation is based on the mechanism of negotiation among intelligent agents (IAs) using a blackboard as a media where each agent exchange information for collaboration. The key members communicate and interact each other in order to generate the solution for efficient material supply process.

- Symeonidis & all (2003) present a MAS for facilitating Supply Chain and Customer Relation (SC-CR systems These system can be viewed as networks of collaborative units that regulate, control and organize all distributed activities involved in procurement, manufacturing, order processing, order transaction and product distribution.

The main objective of their work is the optimization of the quality of services provided by the existing ERP, which provides a robust means for storing and manipulating a large amount of data on company transactions. The choice of developing their architecture as a MAS, provides the advantage of untroubled modification and extension of the system, according to altering company requirements.

## 4. PRODUCT MANAGEMENT AND TECHNOLOGIES

The reduction of product development cycle time and the improvement of product quality has been supported in the last two decades through the implementation of various computer aided technologies. The introduction of the related tools, however, was often only task instead of process oriented. Departments tried to automate their tasks as good as possible and created an environment which best helps to meet their targets. In practice, only little integration took place. Each tool produced its own data format which led to so called 'islands of automation.

This problem could only partially be resolved by various interfaces and data exchange formats.

Romano (2001) in this context states, that during the engineering supply chain, manufactures and suppliers need to work as design partners, however their design and manufacturing systems are frequently incompatible. This introduces very significant time and cost penalties as they attempt to share design information. Moreover, the large amount of data generated by the various tools was often not centrally organised and therefore inaccessible to others. New developments instead of using already designed parts were often the costly consequences.

Realizing this problem and the associated business opportunity several software companies started in the mid 1980s to develop Product Data Management (PDM) systems that initially provided vaulting and file management capabilities for engineering documents. In the late 1980s and early 1990s engineering change management to control and track the changes made to engineering data was added to the functionality together with configuration and classification management capabilities. As a result of design systems and teamwork approaches entering the market PDM systems started to support the management of complex relationships between parts, assemblies, drawings, metadata, people and groups of people.

Since the main functionality of PDM has been design centric data and workflow management it was mainly used in the engineering organisation, sometimes in the manufacturing organisation. This was due to the existing technology which made the accessibility for other than design engineering departments difficult. The software was not usable without extensive training and the files could only be viewed in the native design system.

In the late 1990s a new class of software, Product Lifecycle Management (PLM) systems evolved from PDM systems. Besides the term PLM there are other terms existing that were established by different Research Firms[1]. With the advances in user interfaces and database, viewing as well as the Internet technology the technological prerequisites were provided to share data more easily. Providing integrated visualisation and Digital Mock-Up (DMU) Tools, PLM systems make it possible to view, mark-up and redline native design data without the need for having access to the native design system. The Internet serves as highly effective platform to communicate product data information far beyond the engineering organisation.

Nadamuni (1999) therefore states that the vision for PLM is to do more than just design centric data and workflow management and become an enterprise application that ties together all the information sources in a corporation.

A PLM system can be described as an enterprise-wide Information Technology (IT) infrastructure to support management of product definition throughout its complete lifecycle (from initial concept to product obsolescence). Including workflow management, PLM systems, as a single source of product information, ensure that up to date information are available and accessible for the right people in the right format at the right time. It is also viewed as an effective tool in managing the product definition supply chain by serving as an informational bridge connecting OEM's, partners, subcontractors, vendors, consultants and customers (Miller, 1998).

PLM systems provide a consistent view on product related information in the extended enterprise whereas the easily sharing of product data facilitates real-time collaboration across departmental barriers and among geographically dispersed individuals and groups.

---

[1] CPC (Collaborative Product Commerce), cPDM (collaborative Product Definition Management, ePLM (electronic Product Lifecycle Management), PDC (product definition and commercialisation), PIM (Product Information Management).

Consequently the Aberdeen Group (1999) defines CPC respectively PLM as a class of software and services that uses Internet technologies to permit individuals – no matter what role they have in the commercialisation of a product, no matter what computer based tool they use, no matter where they are located geographically or within the supply net - to collaboratively develop, build and manage products throughout the entire lifecycle.

To cope with the industry as well as social trends and to gain competitive advantage a new integrated approach for developing products with respect to the whole product lifecycle has to be taken. Integration has to take place internally (between different departments) as well as externally (i.e. with suppliers and customers). PLM means the management of comprehensive, accurate and timely information over the entire product lifecycle in order to realise collaborative product development. PLM systems are gaining acceptance for managing all information about a corporation's products throughout their full lifecycle, from conceptualization to operations and disposal (including requirements management, design, engineering, manufacturing, procurement, sourcing, maintenance and service, and more. These sub-processes are linked by a business process workflow). The PLM systems also form the apex of the corporate software hierarchy and as such depend on subsidiary systems for detailed information capture and dissemination. The tools that support PLM philosophy can be broadly categorized as (1) Authoring tools that create product content, and (2) Optimization and decision support system tools that manage and optimize the processes involved in the lifecycle of the product.

It is therefore important not only to take the technological but also the organizational perspective of PLM into account. Collaborating and sharing of data within a company and especially in the extended enterprise challenges existing processes and culture. For that reason, the success of PLM heavily depends on the willingness of the organization to accept change and especially on the people that must form inter-organizational and cross-functional teams to collaboratively develop and manage products.

In a context of product development process that is distributed among many value chain participants, PLM also delivers application support to neatly connect to and accommodate the workings of other processes, including ERPs, CRM, Component Supplier Management (CSM), etc. The benefits of PLM are realized once previously disparate systems such as CAD, PDM, CRM, ERP, SCM.., are integrated (ie. CAD and PDM can share existing design data with the CRM sales configurator). These configurations become the preferred options or option packages available to the customer. The CRM sales configurator may even be powerful enough to present configurations beyond the pre-engineered ones, but within engineered constraints, allowing customers more freedom to choose exactly the combination of features they want. When giving customers this degree of flexibility, SCM also plays a role at the beginning of the development process, enabling a company to "build to order". The customer-selected preference is passed to the SCM requirement planning module. SCM generates and manages the requirements and then passes them to PLM, where they enter the design stage of product development. Parts are sourced using a Component Supplier Management system, also part of SCM. This allows leverage of referred suppliers and design re-use. PLM portal technologies facilitate supplier involvement during the design stage.

Once the initial customer-specific design is proposed, the next step is to simulate the product performance on the computer, saving tremendous time and money over creation and destruction of physical prototypes. And if the simulation shows that the virtual prototype isn't optimized, there's still enough time and money left to go back and try again. Next, manufacturability is verified through advanced simulation, and a final Manufacturing Bill of Material (MBOM) is generated. The MBOM, process plans and other technical data are passed to ERP and SCM for product production.

In fact PLM plays a significant role even while a product is in service. For example, PLM can help companies reduce service and maintenance costs by providing product content information to call centers and field service centers. Call centers can dynamically share a three dimensional model of the product while interacting with a customer. While conducting field service, maintenance engineers can use three dimensional product information to understand maintenance procedures, shortening maintenance times and ensuring higher product quality. As parts are repaired or replaced, the product model can even be updated to reflect the product as maintained, versus the product as built.

### 4.1 Life-cycle standardisation technologies : the PLCS initiative…

PLCS (Product Life-Cycle Systems) is the first and only programme under the auspices of ISO/TC184/SC4 to build upon ISO 10303 (STEP) in extending the capability of STEP to enable the exchange, sharing and archiving of product data in all aspects of though-life support for an extensive range of industry sectors. PLCS initiative aims to accelerate development of new standards for product data in the area of support during operation and disposal of complex engineering assets. The initiative is international and produces draft ISO standards as the result of an initial three-year work programme. These standards are the mechanism by which to ensure support information is aligned with the evolving product definition over the entire life-cycle.

The first year was spent in capturing through-life business process models that represent the essential facets of life-cycle support. The four specific areas for the modelling were:
   • Support Engineering – specifying, providing and sustaining the support infrastructure (the support system);
   • Configuration Management – managing changes to significant product items within the engineered asset and associated support system, including tracking of serial-numbered items;
   • Resource Management – buying, storing, packing, moving, issuing and disposing of product items and elements of the support system;
   • Maintenance and Feedback – maintaining, testing, diagnosing, calibrating, repairing and modifying product items and support elements, including the roles of schedules, resources and feedback.

## 5. CONCLUSIONS

In the new economic context, companies have focused on their core competency and externalised many sorts of activities. This outsourcing of some parts of their production, has led to a wholly distributed environment where performances are reached through highly interconnected organisations and processes. It has also put the stress on the role of integration and coordination to get a lot out of this new kind of business.

It is obvious that the availability of information technology has played a role to this issue in supporting these interactions to implement relationship strategies. IT permits databases consistency, real-time data exchange and, information sharing that are the bases for integrated and collaborative work.

In this paper, we highlighted the importance of software application integration, such as ERP and EAI, as a base for integration and development of collaborative tools and methods. We examined some industrial cross-functional integration through co-managed processes that integrate both suppliers and customers (VMI and CPFR), and show how Multi Agent Systems could help in developing future research dealing with these distributed organisations. Finally, we have drawn attention to the significant role of product information as a vector of integration all along the supply chain.

The future generation of applications will adapt to the internet-based world of today and tomorrow, through changes in functionality, technology and architecture:
- functionality changes will become deeper and more specific to industry domain requirements
- technology will evolve to leverage the internet for both inter-enterprise connectivity and a unification of the end-user experience
- architectural changes will allow for more integration and interoperability.

These changes will also cause vendors and user enterprises to evolve:
- vendors will focus more specifically on domain requirements, providing feature-rich applications that emphasise complete process integration
- enterprises, freed from the monolithic architectures of existing systems will create system deployment strategies that best fit enterprise needs – without relying on a single vendor to deliver all application components
- standards for data exchange will become of increasing importance in facilitating integration and interoperability.

However, it is important to notice that despite the increasing capabilities of the technology, managing and developing collaborative attitude need to be discussed within and between the partners. New supply chain systems process data as they are programmed to do, but the technology cannot absorb a company's history so that cultural impact of Business Process Reengineering for example, is in fact more challenging than the technical issues. The difficulties in organisational transformation should not be underestimated; changing people and the way they do business is, in fact much more difficult than implementing new technologies.